\documentclass[aps,prl,twocolumn,groupedaddress]{revtex4}

\usepackage{graphicx}
\usepackage{dcolumn}
\usepackage{bm}
\usepackage{color}

\begin{document}

\newcommand{\beq}{\begin{equation}}
\newcommand{\eeq}{  \end{equation}}
\newcommand{\bea}{\begin{eqnarray}}
\newcommand{\eea}{  \end{eqnarray}}
\newcommand{\bit}{\begin{itemize}}
\newcommand{\eit}{  \end{itemize}}

\title{Scaling laws in the quantum to classical transition in chaotic systems }

\author{Diego A. Wisniacki}
\affiliation{Departamento de F\'{\i}sica ''J. J. Giambiagi'', FCEN, UBA,
1428 Buenos Aires, Argentina.}

\author{Fabricio Toscano}

\affiliation{Funda\c c\~ao
Centro de Ci\^encias e Educa\c c\~ao Superior  a Dist\^ancia do
Estado do Rio de Janeiro, 20943-001 Rio de Janeiro, RJ, Brazil.}

\affiliation{Instituto de F\'{\i}sica, Universidade Federal do Rio
de Janeiro, Cx. P. 68528,\\
21941-972 Rio de Janeiro, RJ, Brazil.}

\date{\today}

\begin{abstract}
We study the quantum to classical transition in a chaotic system surrounded by 
a diffusive environment. The emergence of classicality 
is  monitored by the Renyi entropy, a measure of the entanglement of a system 
with its environment. We show that the Renyi entropy has a transition from quantum 
to classical behavior that scales with $\hbar^2_{\rm eff}/D$,  where $\hbar_{\rm eff}$ is the
effective Planck constant and $D$ is the strength of the noise.
However, it was recently shown that a different scaling law controls the quantum to classical transition when it is measured comparing the corresponding phase space distributions.
We discuss here the meaning of both scalings in the precise definition of a frontier between 
the classical and quantum behavior. We also show that there are quantum coherences that the  
Renyi entropy is unable to detect which questions its use in the studies of decoherence.

\end{abstract}


\maketitle

Since the birth of the quantum theory, the fundamental question of how a
quantum mechanical system starts to behave classically has been 
the subject of an intense debate. 
In the last years, it has been
well established that decoherence induced by the environment is the main 
ingredient in the quantum-to-classical transition \cite{joos,zurek-1}. However, a 
clear way to describe the fuzzy boundary between the quantum and the classical
world has not been found yet. 

Many attempts have been done to understand the loss of the quantum coherence and several measures of the
restoration of the classicality have been proposed specially in chaotic systems.
Recently, it was conjectured that, in the presence of noise, a single composite parameter controls the quantum-to-classical transition of classically chaotic systems 
\cite{pattanayac-1}. 
This single parameter should appear in the computation of some measure which directly reflects ''the distance''
between quantum and classical evolution as a function of the effective Planck constant $\hbar_{\rm eff}$ 
({\it i.e} the relative size of the Planck constant),   the strength $D$ of the coupling with the environment  and  the Lyapunov coefficient $\lambda$ of the classical dynamics of the system.
Thus, the computed ''distance''  would be controlled by a combined
parameter of the general form $\xi=\hbar_{\rm eff}^{\alpha}\lambda^{\beta}D\gamma$ when the transition from  quantum to  classical behavior occurs.
This conjecture was shown to be valid, for example in  \cite{toscano-davidovich}, where the phase 
space integral 
of the modulus of the difference between the Wigner function and the classical distribution was used as a  measure of the quantum to classical ``distance''.

Another usual quantity to monitor the transition from quantum to classical behavior
is the purity $P=Tr [ \rho^2 ]$, which is a measure of the 
entanglement of the system with the environment.
Any function of the purity has the same information but the so called Renyi entropy, 
$S=-\log(P)$, is particularly interesting because it clearly reveals that above a threshold of the strength
of the coupling with the environment its behavior is dominated by the positive classical Lyapunov exponents of the
system   \cite{zurek-paz,monteoliva-paz} .

In this work we show, for the same system and environment used in \cite{toscano-davidovich}, that  
the behavior of the Renyi entropy $S$ (and thus the purity $P$) 
is controlled by the combined parameter $\chi^{\prime} \equiv \hbar_{\rm eff}^2/D$.
This  scaling was also founded in  \cite{pattanayac-2} using the Duffing  oscillator, 
suggesting an universality of this scaling law.
In our system the emergence of classicality when monitored with the Renyi entropy should correspond to the regime
$\chi^{\prime}  < 100$.
However, in the regime $1<\chi^{\prime}<100$
the evolved Wigner function displays an interference fringe pattern that clearly indicates that not all  the quantum  
coherences of the system have dissapeared.
We explain that while $\chi^{\prime}$ controls which quantum terms effectively contribute  
in the master
equation for evolution of the Wigner function, the fine coherent structure of this 
function is controlled by the composite parameter $\chi= \hbar_{\rm eff}^2K/4D^{3/2}$ \cite{toscano-davidovich}.
We also show that only in this essentially classical regime  and for short times,  
the behavior of the Renyi entropy is independent on the diffusion constant and dominated, in our case, 
by a classical Lyapunov exponent. 

\begin{figure}{h}
\setlength{\unitlength}{1cm}
\begin{center}
\includegraphics[width=6.cm,angle=-90]{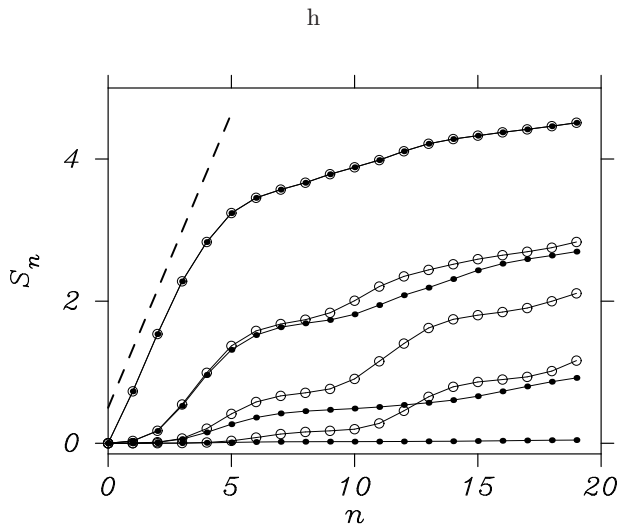}
\end{center}
\caption{Quantum ($\bullet$) and classical ($\circ$) Renyi entropy as a function of the number of the kicks for the KHO with $K=2$ and $\eta=\sqrt{\hbar_{\rm eff}/2}=0.125$. From bottom to top, the diffusion constants are 
$D= 2.56 \times 10^{-7},
6.4 \times 10^{-6}, 9\times 10^{-5}$ and $ 4\times 10^{-3}$. The dashed line is a linear function with a slope given 
by the logarithm of the classical expansion coefficient at $(q,p)=(0,0)$. }
\label{fig1}
\end{figure}

The system we analyze is the kicked harmonic oscillator (KHO) whose quantum Hamiltonian is 
\begin{equation}
\label{Hamiltonian_kicked_rotor}
H=\frac{\hat{P}^2}{2m}+\frac{1}{2}m\nu^2\hat{Q}^2+
A\;\cos(k \hat{Q})\;\delta_n(\tau),
\end{equation}
where $\delta_n(\tau)\equiv\sum_{n=0}^\infty \delta(t-n\tau)$, 
$m$ is the particle mass, $\nu$  the oscillator frequency, $A$ the amplitude of the kicks
and $\tau$ the interval between two consecutive kicks. 
Working with dimensionless quantities $\hat{q}=k\hat{Q}$, $\hat{p}=k\hat{P}/m\nu$ and $K=k^2A/m\nu$,
we can define an effective Planck constant $\hbar_{\rm eff}$, so 
that $[\hat{q},\hat{p}]=2i\eta^2\equiv i\hbar_{\rm eff}$, with the so called Lamb-Dicke parameter
$\eta=k\Delta Q_0=k\sqrt{\hbar/2m\nu}$  being the quotient between the width $\Delta Q_0$ of the
ground state of the harmonic oscillator and the wavelength $\tilde{\lambda}=2\pi/k$ of the position dependent pulse.
The classical dynamics in
phase space of the KHO is unbounded 
and the stroboscopic sections exhibit a mixed dynamics
with stable islands surrounded by a chaotic sea  \cite{ZaslavskySagdeev}.
Here, we consider the relation $\tau=T/6$ between the period of the harmonic 
oscillator $T=2\pi/\nu$ and the interval between kicks $\tau$. Thus, the classical phase space structure
corresponds to a stochastic web, its thickness governed by $K$. We have considered
$K=2$, thus the phase space origin  is an hyperbolic fixed point and 
the area around it shows an essentially
chaotic behavior \cite{toscano-wisniacki}.

In order to study the decoherence effects that induce a quantum-to-classical transition
we coupled the KHO,  as usual \cite{pattanayac-1,pattanayac-2,monteoliva-paz,toscano-davidovich}, to 
a thermal reservoir with average population $\bar{n}$  and coupling constant $\Gamma$
in the Markovian and weak coupling limit, and we considered  the purely diffusive regime ({\it i.e.} 
the high temperature regime $\bar{n} \rightarrow \infty $ together with $\Gamma \rightarrow  0$ and
$\tilde{\Gamma}\equiv \bar{n}\Gamma\eta^2$ a constant value). 
For this kind of environment the master equation for the evolution of the Wigner function $W({\bf x})$ reads
\begin{eqnarray}
\frac{\partial W}{\partial t}&=&L_{cl}+L_q+T=\{H,W\}+\nonumber\\
&+&\sum_{m=1}^{\infty}\frac{(-1)^m\hbar_{\rm eff}^{2m}}{2^{2m}(2m+1)!}\frac{\partial^{2m+1}V}{\partial q^{2m+1}}
\frac{\partial^{2m+1}W}{\partial p^{2m+1}}+\nonumber \\
&+&\tilde{\Gamma}\left(\frac{\partial^2 W}{\partial q^2}+\frac{\partial^2 W}{\partial p^2}\right)\;.
\label{master-equation}
\end{eqnarray}
The equivalent evolution equation for the classical distribution only contains the 
evolution generated by the Poisson bracket $L_{cl}$ and the term $T$ that comes from the coupling with the diffusive
environment. The quantal term $L_q$ comes from the nonlinear part of the Hamiltonian that in our case
are simple the kicks $V(q)=K\cos(q)\delta_n(\tau)$. 
The strength  of the interaction with the enviroment is controled by the dimensionless diffusion constant, $D\equiv\tilde{\Gamma}\tau$.

Let us first consider the quantum-to-classical transition by monitoring the behavior
of the classical and quantum Renyi entropies $S=-\log(P)=-\log(2 \pi \hbar_{\rm eff} \int W^2({\bf x}) d{\bf x})$ ($W$ is the Wigner function or the
classical phase space distribution).
In Fig. \ref{fig1} we show the behavior of the entropies in the KHO with $K=2$ and 
$\eta=\sqrt{\hbar_{\rm eff}/2}=0.125$ for several values of the  strength of the interaction with the enviroment $D$. 
The initial state is a coherent state centered at the origin of phase space with width 
$\Delta q=\Delta p=\eta$.
We can see that for extremely small values of $D$
($D=2.56 \times 10^{-7}$), and short times, the quantum and classical evolution are nearly unitary, and thus the Renyi entropies are $S_n \approx 0$. Increasing the value of the diffusion constant $D$ and in the regime where the initially localized wave packet stretches in the direction of the unstable classical manifold
the quantum and classical Renyi entropy exhibit different time regimes [Fig. \ref{fig1}].
When $D$ is still relative small, we can see two regimes \cite{toscano-wisniacki-2}.
In the first one the behavior of both entropies is essentially quadratic in time and the differences between them
are negligible.
Then, there is a linear regime whith a slope that depends on $D$ and can be different for the quantum and the classical case. 
This linear  regime occurs for times of the order of the well known 
log-time $t_E=\log(1/\hbar_{\rm eff})/(2 \lambda)$, with 
$\lambda$ the Lyapunov exponent or, as in our example, the logarithm of the expansion
coefficient of the linearization of the dynamics at the origin (hyperbolic fixed point)
\cite{toscano-davidovich}. 
Finally, we can see that  for sufficiently greater values of $D$ the linear growth of the quantum and the classical Renyi entropy
are identical and given by $\lambda$ \cite{zurek-paz}.

\begin{figure}
\setlength{\unitlength}{1cm}
\begin{center}
\includegraphics[width=8cm,,angle=0]{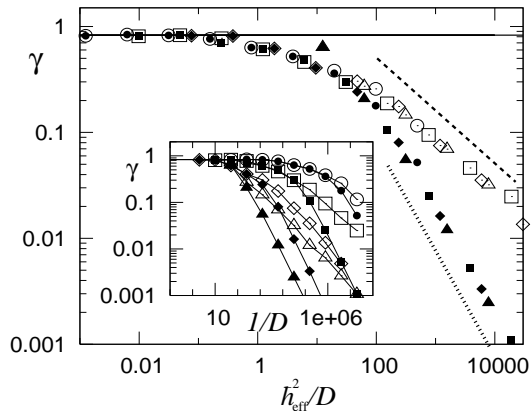}
\end{center}
\caption{Slope $\gamma$ of the linear growth of the  quantum (full symbols) and classical (empty symbols) Renyi entropy as a function of  $\hbar_{\rm eff}^2/D$ (main plot) and $1/D$ (inset) ($D$ is  the diffusion constant) 
for $\eta=\sqrt{\hbar_{\rm eff}/2}=0.5$ (triangles), $0.3125$ (diamonds), $0.125$ (squares) and
$0.0625$ (circles). We have also plotted the functions $\gamma \propto D/\hbar_{\rm eff}^2$
(dotted line) and $\gamma \propto \sqrt{D}/\hbar_{\rm eff}$
(dashed line). See text for details. }
\label{fig2}
\end{figure}

\begin{figure}
\setlength{\unitlength}{1cm}
\begin{center}
\includegraphics[width=7.5cm,,angle=0]{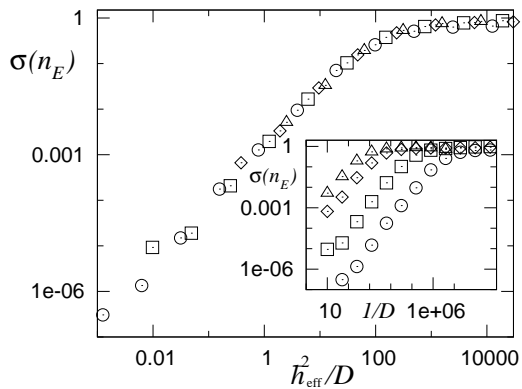}
\end{center}
\caption{ The measure $\sigma(n_E)$ (see text) of the differences between  the quantum and classical  Renyi entropy.
The initial state is a coherent state centered around the origin of phase space with width $\Delta q=\Delta p=\eta$,
with $\eta=\sqrt{2 \hbar_{\rm eff}}=0.5$ (triangles), $0.3125$ (diamonts), $0.125$ (squares) and
$0.0625$ (circles). For each $\hbar_{\rm eff}$ the Renyi entropies were computed at the log-time $t_E\approx n_E\tau$. }
\label{fig3}
\end{figure}

We have studied systematically the dependence of
the linear growth regime with the diffusion constant $D$ and $\hbar_{\rm eff}$. 
In Fig. \ref{fig2} (inset)  we show
the slope $\gamma$ for the linear growth of the  quantum and classical Renyi entropy
as a function of the diffusion constant $D$ for several values of $\hbar_{\rm eff}$.    
Both slopes $\gamma$ approach $\lambda$ (horizontal line) for all the $\hbar_{\rm eff}$ considered 
but at different values of the diffusion constant $D$.
However, when $\gamma$ is plotted as a function of $\hbar_{\rm eff}^2/D$ a clear scaling emerges.
We can see that the classical and quantum entropy linear growth rate measured by $\gamma$ are equal for 
$\hbar_{\rm eff}^2/D < 100$, but it is only in the range $\hbar_{\rm eff}^2/D<1$ that the classical and quantum
slope approach $\lambda$ (the so called ``Lyapunov'' regime). 
On the other hand, for  $\hbar_{\rm eff}^2/D > 100$
the linear growth rate of the Renyi entropy depends linearly with the diffusion constant $D$. This is the
so called Fermi golden-rule regime that was explained in Refs. \cite{zurek-paz, 
cucchietti-paz}. For the classical case, we found that the Renyi entropy linear growth rate 
scales with $\sqrt{D}$.
This fact deserves further investigation \cite{toscano-wisniacki-2}.
The scaling with $\hbar_{\rm eff}^2/D$  can also be appreciated in Fig.(\ref{fig3}) where we plot the measure,
$\sigma(n)\equiv |S_q(n)-S_{cl}(n)|/S_q(n)$, of the ``distance'' between the quantum Renyi entropy, 
$ S_q$, and the classical one, $S_{cl}$, computed at the number of kicks  
$n=n_E \approx t_E/\tau=\log(1/\hbar_{\rm eff})/(2\tau \lambda)$.
 

\begin{figure}[h]
\setlength{\unitlength}{1cm}
\begin{center}
\includegraphics[width=7.5cm,,angle=0]{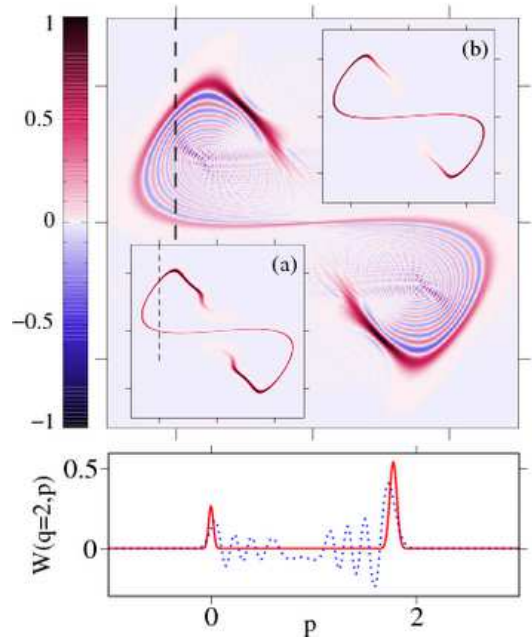}
\end{center}
\caption{(Color online) Wigner distribution (top panel) and classical distribution [inset (a)] inmediatly before the kick $n=7$ for  $\eta=\sqrt{2 \hbar_{\rm eff}}=0.125$ and difussion constant  $D=9\times 10^{-5}$. Inset (b) Wigner distribution inmediatly before the kick $n=8$ for  $\eta=\sqrt{2 \hbar_{\rm eff}}=0.0256$ and difussion constant  $D=9\times 10^{-5}$. Bottom panel:
Wigner distribution (blue dotted line) and classical distribution (red solid line) evaluated on (q=-2,p) (dashed lines in the top panel).}
\label{fig4}
\end{figure}

Another approach to monitor the emergency of  classicality is to directly compare 
the evolution of the Wigner function and the 
corresponding classical distribution in the presence of the environment
 \cite{toscano-davidovich,toscano-wisniacki}.
 A different scaling law appears in this approach where it was shown that
 the Wigner function and the classical distribution are essentially equal in the parameters regime
 $\hbar_{\rm eff}^2/D<<1$ and $\chi= \hbar_{\rm eff}^2K/4D^{3/2} \lesssim 1$. 
 However, we have seen that the linear growth of the classical and quantum Renyi entropy are 
 equal for  $\hbar_{\rm eff}^2/D<100$. Even more, from Fig. (\ref{fig1}) we see that the differences 
between the quantum  and classical Renyi entropy are negligible beyond the linear regime up to
$n<9$ for example for $D=9\times 10^{-5}$ where $\chi^{\prime}=\hbar_{\rm eff}^2/D\approx 10 <100$ but 
$\chi= \hbar_{\rm eff}^2K/4D^{3/2}\approx 571 >1$.
In Fig. \ref{fig4}, we show the quantum and classical distribution immediately before
the kick $n=7$ for $\chi^{\prime}\approx 10$ where we clearly
see that the Wigner distribution shows important interference  fringes arising
from quantum coherences not detected by the Renyi entropy. 
These quantum coherences separate the quantum from the classical behavior.

We can understand the different role of the scalings $\chi^{\prime}$ and $\chi$ from 
the exact solution of Eq. (\ref{master-equation}) between kicks.  
This solution can be written as the map: 
%
$W_{n+1}({\bf x})=
\int d{\bf x'}\;\tilde{L}({\bf x}^R,{\bf x'})\;W_n({\bf x'})$
where $W_{n+1}$ and $W_n$ are the Wigner functions or the classical phase space distributions immediately before
the kicks $n+1$ and $n$ respectively, $\tilde{L}$ is the quantum or classical propagator
of one kick plus the action of the diffusion reservoir and
${\bf x}^R\equiv(q^R,p^R)\equiv{\bf R}^{-1}({\bf x})$ correspond to the harmonic evolution
\cite{toscano-davidovich,toscano-wisniacki}.
It is instructive to write the one step quantum nonunitary propagator 
$\tilde{L}$ for an arbitrary position dependent kick $V(q)$:
\begin{eqnarray}
\label{smoothed-quantum-propagator}
\tilde{L}({\bf x}^R,{\bf x'})&\equiv&
 \frac{e^{-\frac{\left(q^R-q'\right)^2}{4D}}}{\sqrt{4\pi D}}
\int_{-\infty}^{+\infty}
\frac{d\mu}{2\pi \hbar_{\rm eff}}
e^{-\frac{D\mu^2}{\hbar_{\rm eff}^2}}\;
\times  \nonumber\\
& \times&\;\;e^{\frac{i}{\hbar_{\rm eff}}\left[V(q'-\mu/2)-V(q'+\mu/2)-
\mu(p^R-p')\right]},
\end{eqnarray}
(for $V(q)=K\cos(q)$ this formula  corrects for a small mistake in Eq. (16) of Ref. \cite{toscano-wisniacki} ).
If we write,
\begin{equation}
\label{difference-of-V}
V(q'_{-})-V(q_{+})=- \mu \partial_q V
-\sum_{m=1}^{\infty}
\frac{\mu^{2m+1}}{2^{2m}(2m+1)!}\partial^{2m+1}_q V\,,
\end{equation}
where $q'_{\pm}=q'\pm \mu/2$ and $\partial_q ^{2m+1}V\equiv \partial^{2m+1}V(q^{'})/\partial q^{2m+1}$ ($m=0,1,\ldots$)
\footnote{Note that
for the KHO we have   $\partial_q^{2m+1}V(q')=(-1)^{m-1}\sin(q)$, thus we get 
the simplification $V(q'_{-})-V(q_{+})=2\sin(q')\sum_{m=0}^{\infty}(-1)^m(\mu/2)^{2m+1}/(2m+1)!=
2\sin(q')\sin(\mu/2)$.}, and 
we kept only the first term we recover the classical propagator of the classical distribution:
$\tilde{L}^{cl}({\bf x}^R,{\bf x'})=e^{-\left(x^2+y^2\right)}/4\pi D$
%
where  $y\equiv\left( p^R- p^{'}  + \partial_q V \right)/2\sqrt D$ and $x\equiv \left( {q^R - q^{} } \right)/2\sqrt D$. 
Thus, we see from Eq.(\ref{difference-of-V}) that the same high order odd derivatives that enter 
in the quantal ``correction''
$L_q$ in the master equation (\ref{master-equation}) enter in the one step propagator (\ref{smoothed-quantum-propagator}).
Therefore, the value of the  variance $\chi^{\prime}/2=\hbar_{\rm eff}^2/2D$ of the  
Gaussian factor $\exp(-D\mu^2/\hbar_{\rm eff}^2)$ in Eq.(\ref{smoothed-quantum-propagator})
is crucial in the quantum-to-classical transition because it controls which terms of this quantal ``correction''
effectively contribute. Thus, if the variance $\chi^{\prime}/2$ is sufficiently small for the quantal ``corrections'' we have
$|\mu|^{2m+1}|\partial^{2m+1}_qV|/\hbar_{\rm eff}c< |\mu|^3 |\partial V_{\rm max}|/\hbar_{\rm eff}c$ where
$c\equiv 2^{2m}(2m+1)!$ and $|\partial^{2m+1}_qV|<|\partial V_{\rm max}|$ is  an upper bound valid in the region where
the initial wave packet spread (for the KHO, $|\partial V_{\rm max}|=K$ in all position space).
But, $|\mu|^3 |\partial V_{\rm max}|/\hbar_{\rm eff}c$ it is  only  a small correction iff 
$\chi=\hbar_{\rm eff}^2|\partial V_{\rm max}|/4D^{3/2}<<1$ where, following the steps described in \cite{toscano-wisniacki},
we can write
\begin{equation}
\label{approx_smooth_quantum_propagator}
L ({\bf x}^R,{\bf x'}) \approx L^{cl}({\bf x}^R,{\bf x'})\left[1  + \chi 
\frac{\partial^3_qV(q^{'})}{|\partial V_{\rm max}|}\;f(y) \right]\,,
\end{equation}
with $f\left( y \right) = {1}/{4}\left( y - {2y^3 }/{3} \right)$.
Based on this result, in Ref. \cite{toscano-davidovich,toscano-wisniacki} 
it is showed that the distance between the corresponding phase-space distributions,${\mathcal D}_n\equiv
\int d{\bf x}\left|W_{n}({\bf x})-W_{n}^{cl}({\bf x})\right|\;,$  that quantify the quantum effects, scales with the single parameter $\chi$.
In Fig. (\ref{fig4}) (inset b) we can see that for $\chi=1$ the evolved Wigner distribution 
does not display essentially any quantum coherence.

We have shown that the  quantum-to-classical transition in the presence of
a diffusive
enviroment is governed by the single parameter $\chi'=\hbar^2/D$ when it
is measured with
the Renyi entropy.
Also, we have shown that $\chi^{\prime}$ controls whose terms are the
leading order quantum corrections in the master equation for the Wigner
function.
However, for a class of kicked systems
we have shown that the combine parameter that indicates
in which regime the leading correction is negligible
is $\chi=\hbar_{\rm eff}^2|\partial V_{max}|/4D^{3/2}$
(that for the KHO reduce to $\chi= \hbar_{\rm
eff}^2K/4D^{3/2}$).
It was previously shown that the distance between
the classical and quantum phase space distributions, which measure the
quantum signatures, is proportional to the parameter $\chi= \hbar_{\rm
eff}^2K/4D^{3/2}$.
Thus, it is the combined $\chi$ parameter that defines the precise
frontier between the classical
and quantum behavior of a system in a purely diffusive reservoir.
Furthermore, our results clearly indicate that the dynamics of the Renyi
entropy or the purity is unable to detect not negligible quantum coherences
and  therefore questioned its use to study the
quantum to classical transition.


We acknowledge F. Lombardo, J. P. Paz, M. Saraceno, R. Vallejos and  E. Vergini for their fruitful comments.
DAW acknowledge support from CONICET (PIP-6137) and UBACyT (X237).

\end{document}